# A SURVEY ON DATA AND TRANSACTION MANAGEMENT IN MOBILE DATABASES

D. Roselin Selvarani [1] and Dr. T. N. Ravi[2]

[1]Department of Computer Science, Holy Cross College, Bharathidasan University,
Tamil Nadu, India.
*jmjroselin@yahoo.co.in*
[2]Department of Computer Science, Periyar E.V.R. College, Bharathidasan University,
Tamil Nadu, India.
*proftnravi@ymail.com*

## ABSTRACT

*The popularity of the Mobile Database is increasing day by day as people need information even on the move in the fast changing world. This database technology permits employees using mobile devices to connect to their corporate networks, hoard the needed data, work in the disconnected mode and reconnect to the network to synchronize with the corporate database. In this scenario, the data is being moved closer to the applications in order to improve the performance and autonomy. This leads to many interesting problems in mobile database research and Mobile Database has become a fertile land for many researchers. In this paper a survey is presented on data and Transaction management in Mobile Databases from the year 2000 onwards. The survey focuses on the complete study on the various types of Architectures used in Mobile databases and Mobile Transaction Models. It also addresses the data management issues namely Replication and Caching strategies and the transaction management functionalities such as Concurrency Control and Commit protocols, Synchronization, Query Processing, Recovery and Security. It also provides Research Directions in Mobile databases.*

## 1. INTRODUCTION

The usage of mobile database has increased rapidly because of the continuous growth of the Hardware devices with greater storage capacity and more powered CPU along with the fast development of the Wireless technology. Mobile devices are gradually more used for database driven applications like Sales Order Entry, Product Inventory Tracking and Customer Relationship Management. The way in which mobile applications access the data and manage them is changed completely due to these applications. In these applications data are moved closer to them to improve the efficiency and autonomy instead of storing them in a central database. This new style creates many motivating issues in mobile database research. In this paper we survey the issues related only to the data and transaction management and briefly present the state-of-the-art of data and transaction management in Mobile databases.

There are various definitions found in the literature on Mobile Databases. Some authors defined Mobile Database as a database that is stored on the mobile devices such as laptops, PDAs and Cell phones [1,2,3,4,5]. Few other authors stated that it is a distributed database in which the accessing mode is mobile [6,7]. Some other authors described Mobile database as the union of distributed database, disconnected database, Ad-hoc database and broadcast disk. The distributed database is treated as the home for mobile database and the others deal with the access of mobile users [8].

The remaining part of this paper is organized as follows. Section 2 reviews Architecture of Mobile Databases. Section 3 analyses the various Mobile Transaction Models used in the literature so far. Section 4 discusses the existing algorithms / protocols / methods / techniques used in Data and Transaction Management in Mobile Database. Section 5 presents the Issues and Research Directions of Mobile Databases and Section 6 Concludes the paper.

## 2. ARCHITECTURE OF MOBILE DATABASE

### 2.1. Reference Architecture

Vijay Kumar presents Reference architecture (Fig.1) for mobile database system in [7]. In this architecture, Fixed Host (FH) and Base Station (BS) are interconnected through a high speed wired network. One or more BS is connected with Base Station Controller (BSC), which coordinates the operation of BS when commanded by the Mobile Switching Centre (MSC). BSs are incorporated with some simple data processing capability so that they can coordinate with database servers (DBS). Unlimited mobility in PCS and GSM is supported by wireless link between BS and Mobile units. To incorporate full database functionality, it is necessary to incorporate DBSs to PCM or GSM. Each DBS can be reached by any BS or FH. The set of MSC and PSTN (Public Switched Telephone Network) connects the MDS to the outside world.

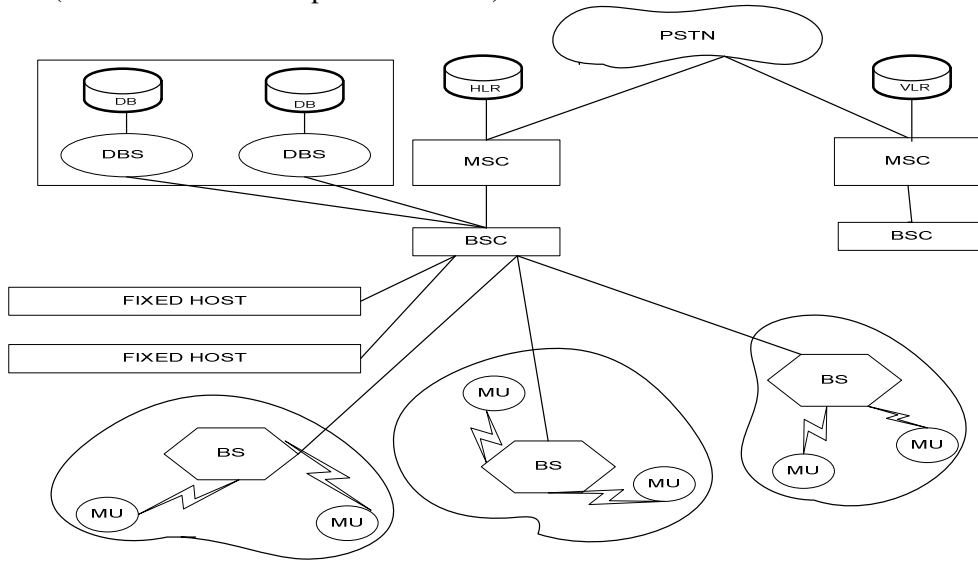

Figure 1. Reference Architecture

## 2.2. Types of Architecture

Apart from the Reference Architecture for Mobile database, there are three types of architectures discussed in the literature namely Client/Server architecture, Server/Agent/Client architecture and Peer-Peer architecture.

### 2.2.1. Client-Server architecture

Figure 2 represents a Client-Server architecture which consists of a server, fixed hosts and mobile clients [8, 9]. A mobile client is in disconnected state when it enters the shadow area of wireless network. In order to save power and expensive wireless communication cost and provide faster response time, it may want to execute transactions in disconnected state intentionally. A mobile host has mobile database hoarded from the server to process transaction locally during disconnection. Whenever wireless connectivity improves or a wired link is available, the mobile client connects with the server and incorporates effects of the transaction committed during disconnection into the server database.

In [10], Publication/Subscription framework is addressed, which can also be considered as a Client - Server architecture. In this framework, the Publisher object acts as a Server whereas the Subscriber object acts as a Client. The authors used this model for Object- Oriented data synchronization of mobile databases over mobile ad-hoc network. In [11], 3-tier replication architecture (Mobile client – Synchronization Server – Database Server) is presented. This model uses a double timestamp strategy to improve the efficiency of synchronization problem in embedded database based Mobile Geospatial Information Service. Also it ensures serializability and consistency of the mobile database system through transaction-level conflict reconciliation.

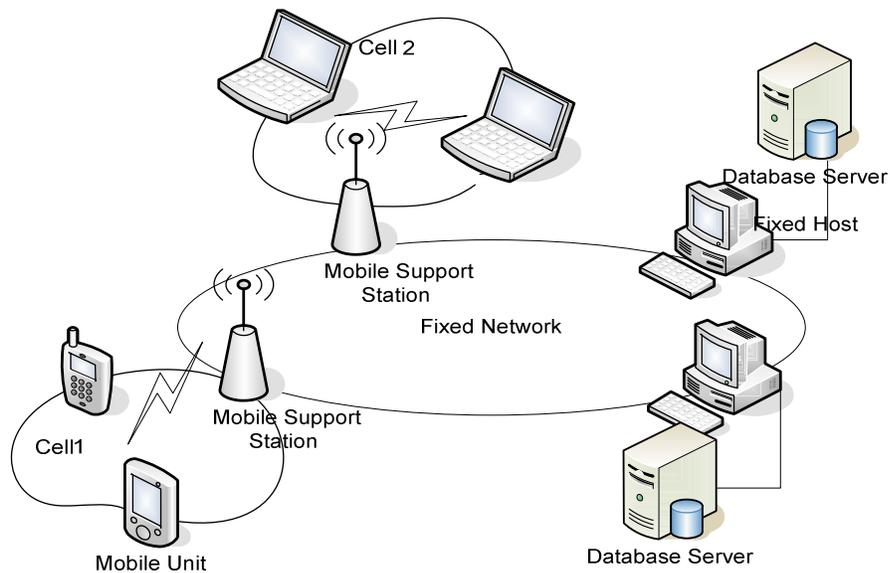

Figure 2.  Client-Server Architecture

Depending on the characteristics of Mobile client, it may be varied from thin to full client that leads to 3 more architecture:  **Thin Client architecture, Full Client architecture and Flexible**

**Client-Server architecture**. In thin Client architecture clients need to run operations on servers. As the resources are restricted in thin client, Server is responsible for all computations. In Full Client architecture, the client should have sufficient resources so that it can do all the functionalities of the server in a disconnected mode. In Flexible Client-Server architecture, the responsibilities of clients and servers as well as the application functionalities can be dynamically relocated. The distinction between clients and servers may be temporarily indistinct for performance and availability purposes. Another related architecture is Push-based data delivery model or **Data Broadcasting Model.** In this model, the Server broadcasts the data on the broadcast channel and the mobile tune to that particular channel to retrieve the information. Several research works are going on in this model but Broadcast related topics are out of the scope of this paper.

### 2.2.2. Server/Agent/Client architecture

Figure 3 depicts a Server/Agent/Client architecture, where an agent is included either in the server or in the client side. An agent is a computer program that can achieve a series of goals of designers and users, and can freely and autonomously operate in the mobile environment. The architecture of mobile database based on agent has three layers: Mobile terminal layer, Mobile agent layer and Server layer. Each layer has some agents and each agent is responsible for a specific task [12].

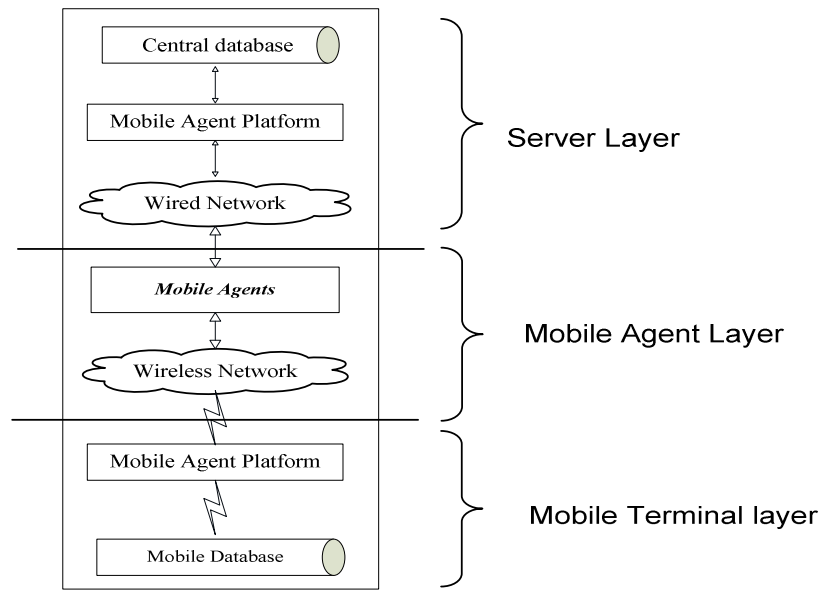

Figure 3. Server/Agent/Client Architecture

### 2.2.3. Peer-to-Peer architecture

The third type of architecture is a Peer-to-Peer architecture (Fig.4). An Ad-hoc network may lead to such type of architecture, in which clients may also communicate with other clients to share their data. The advantage of this type of architecture is a higher degree of availability of data but may compromise consistency, if the clients are allowed to update any data [10].

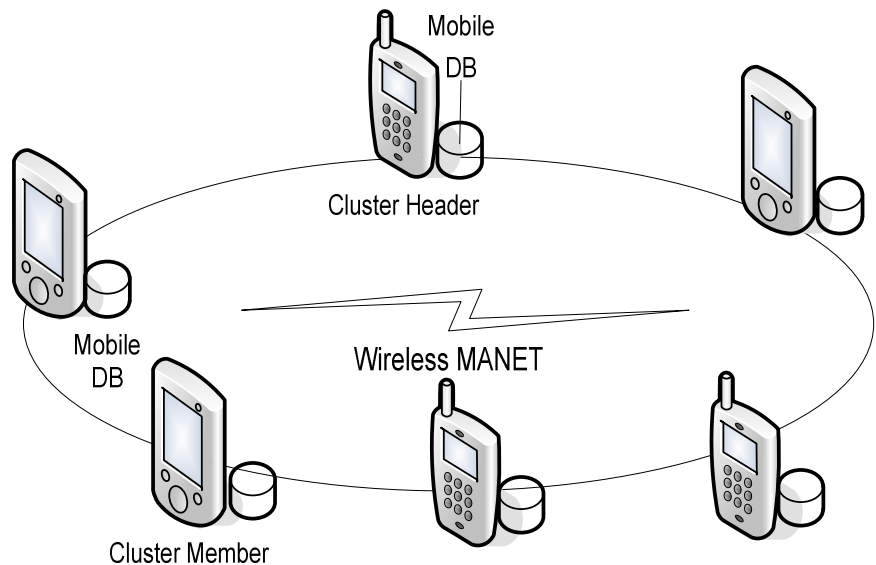

Figure  4.  Peer to Peer architecture (MANET)

## 3. MOBILE TRANSACTION MODELS

Collection of operations that form a single logical unit of work is called a transaction [8]. If it takes place in the mobile environment, it is called Mobile Transaction. In  [7], the characteristics of the Mobile Transactions is provided as follows:

✓ The Mobile transactions are long-lived transactions due to the mobility of both data and users and due to the frequent disconnection.
✓ The mobile transactions might have to split their computations into sets of operations, some of which execute on MH and others on MSS. A mobile transaction shares their states and partial results with other transactions due to disconnection and mobility.
✓ The mobile transactions require computations and communications to be supported by mobile service stations (MSS).
✓ As the MHs move from one cell to another, the states of transaction, states of accessed data objects, and the location information also moves.
✓ The mobile transactions should support and handle concurrency, recovery, disconnections and mutual consistency of the replicated data objects.

Transaction in MDS is executed in a distributed manner (i.e. Client-Server) which may be subjected to further restriction due to inherent constraints such as limited bandwidth. Distributed execution of transaction in the presence of mobility makes transaction commitment and termination quite complex. Mobile transaction may have to suffer "forced wait" or "forced abort" because of the lack of wireless channels (uplink or downlink), and further it may be delayed due to random hand-offs. In addition, a mobile transaction at MH may not be able to complete its execution due to non-availability of complete DBMS capability.

Transaction support is critical in mobile environment. Due to the limitations of mobile environments such as variable bandwidth, disconnections, limited resources on mobile hosts make traditional transaction management techniques no longer appropriate. Several models for mobile transactions have been proposed but it is difficult to have an overview of all of them.

Mobile host (MH) can operate in three different modes: fully connected, disconnected and partially connected. Another mode is a doze mode. MH can decide to switch to a disconnected mode to minimize connection expenses or because no connection is possible. MH can change to a doze mode to save energy. In this mode the clock speed is reduced and no user computation is performed. The MH returns to normal operation upon the receipt of any message.

Transaction Models found in the literature can be classified into three groups based on the locations where the transactions are initiated and executed. A  Mobile transaction may be Initiated as well as Executed by MH, Initiated at MH but Executed by FH and Initiated at MH and Executed among MH and FH depends on the requirements.

## 3.1. Initiated as well as Executed by MH

When a mobile transaction is initiated as well as executed by MH, the MH gets full autonomy. The MH should have relevant data replicated from the central server and capable of doing a local transaction. After completing the transaction, the data in the MH is integrated to the central server. This type of execution model permits the MH to work even in the disconnected mode. For example, consider a transaction carried out by the Mobile Worker. The general procedure is he stores all the required data related to his task on his MH, works independently at the remote site and after the successful completion of his work; he updates the data to the central server. The transaction models that support this type of execution scenario are Pre-Write transaction model [13], and High Commit Mobile(HICOMO) transaction model [14]. **Pre-Write** transaction model proposes two variants of data, pre-write and write to increase the data availability on MHs. **HICOMO** transaction uses Aggregate table for disconnected mode.

## 3.2. Initiated at MH but Executed by FH

When a mobile transaction is initiated by a MH but executed fully on FHs, the MH need not have any data processing capabilities. In this model, the MH sends the query and the data server executes the query and sends the results back to the MH. This type of transaction is more suitable for Location dependent queries. Example transaction models for this type of scenario are Moflex Transaction Model [15] and Pre-Serialization Transaction model [16].

**Moflex** transaction model is based on set of dependencies acceptable goals and rules. It supports location dependent sub transactions**. Pre-serialization** allows site transactions to commit independently of the global transaction. It allows releasing local resources in a timely fashion.

## 3.3. Initiated at MH and Executed among MH and FH

When a mobile transaction is executed between MH(s) and the database server(s) or FHs, MH also should have the server capabilities and minimum communications with the server during the transaction execution. The transaction models supporting this type of execution is Multiversion Transaction Model [17]. To overcome the problem of data access delay in remote transaction, Kam-Yiu Lam et al. propose this model. In this model, a real-time transaction may access stale data at fixed host, provided that they are relatively consistent with the data accessed by the transaction, and the staleness of the data is within the requirements.

In [18], the mobile transaction is initiated at MH and executed in five different scenarios: Entirely at database servers, entirely at the mobile host, at one mobile host and several DB, at several mobile hosts and finally at several mobile hosts and databases are explained. Whereas in [19], only three types of scenarios are discussed for databases on mobile clients: They are Mobile clients which are connected to a wired network, Wireless Network of mobile clients with single-hop distance to each other and Wireless Network of mobile clients some of which are multi-hop distance. The authors also addressed some of the issues for each scenario. In [20], the authors discuss a few mobile transaction models and compare them based on ACID properties.

**Twin-Transaction Model** [21] defines a transaction execution mechanism which satisfies the need of both connected and disconnected modes of operation. A defined resynchronization mechanism achieves a consistent state on reconnection of the mobile host. Zhengwei et al., in [22], propose a novel theoretical mobile web transaction model called **PMTM** (P system-based Mobile Transaction Model) to formalize the behavior of mobile transactions. This model has two transition rules namely Membrane rules and Object rules. The Object rule describes the transitions in membranes whereas the Membrane rule defines the structural modification of membranes.

In [23], Petricia et al. propose an **Adaptable Mobile Transaction Model** which permits defining transactions with several execution alternatives associated to a particular context. The aim of this model is to adapt transaction execution to context variations. In this model Atomicity and Isolation properties are relaxed but conflict serializability is preserved. The advantage of this model is that it improves the commit possibilities and permits to select the way transactions will be executed according to their costs.

Osman Mohammed Hegazy et al. in [24] discuss a new enhanced **shadow paging technique** called a Mobile-Shadow technique for handling mobile transaction processing and disconnection. M-Shadow uses a notation of actionability, which differentiates the actions to be taken during the transaction's validation phase according to the types of affected attributes.

A new transaction scheme called **Surrogate Object Based Mobile Transaction Model** (an Improved Kangaroo Transaction Model) is presented by Ravimaran.S et al., in [25]. The main focus is to support data caching at surrogate object for faster data access and database operations among mobile transactions at different mobile hosts in mobile environment. The experimental results prove that there is a significant reduction in wireless access and abort probability can be obtained with the proposed model. In [26], Tome Dimovski et al. propose a **Connection Fault-Tolerant Model** for mobile environment which reduces the blocking time of resources at the fixed devices provides fast recovery from connection failures owing to mobility of mobile devices and increases the number of committed mobile transactions.

## 4. DATA AND TRANSACTION MANAGEMENT IN MOBILE DATABASES

This section provides an elaborate study on Replication and Caching strategies under data management and the functionalities of Mobile Transaction such as Concurrency Control and Commit Protocols, Synchronization, Query Processing, Recovery and Security under Transaction Management.

### 4.1. Concurrency Control and Commit Protocols

Many valuable attempts were made to implement the concurrency control strategies in mobile environment. Three types of methods were found in the literature for concurrency control mechanism such as Locking, Timestamps and Optimistic Concurrency Control. Even though these methods are well suited for traditional database applications, they are not suitable for mobile environments because of its inherent nature. Transactions initiated by a mobile client that disconnects for longer time period may lead to an unacceptable long locking which may decrease the throughput [27]. When two mobile clients access a data item concurrently where one client tries to read the data item while the other tries to write upon it, it may result in inconsistency. For this purpose a two phase locking protocol is used, which requests the server to lock all the data items demanded. However, this protocol requires the clients to communicate continuously with the server to obtain the necessary locks and detect the data conflicts, and hence is not suitable to the wireless environment where the capacity of communication bandwidth is highly variable and unexpected disconnection may occur. Timeout based approach is proposed in [28] to avoid the problem of starvation due to locking.

 A Timeout based Mobile Transaction Commitment Protocol uses timeouts to provide non-blocking protocol with restrained communication. It faces the problem of the time lag between local and global commit. It is not possible for the mobile clients to execute the transaction within the specified time period, because of the variation in bandwidth disconnection etc. Hence a dynamic timer adjustment strategy is proposed in [29]. In this mechanism the communication overhead is reduced and therefore the transaction throughput is improved. But it suffers from the problem of frequent rollbacks due to regular expiry of the timer. To reduce the frequent rollbacks, a pre-emptive dynamic timer adjustment strategy and a predictive strategy [30] is proposed. As the offline processing capability for individual mobile host may vary, it may execute the transaction faster even if it has requested for execution quite later. For this reason, an optimistic concurrency control technique is frequently used in wireless environments [31]. In [32], an optimistic concurrency control strategy using **on-demand multicasting** is proposed that guarantees consistency and introduces application-specific conflict detection and resolution strategies. The experimental result shows that there is an increase in system throughput by reducing the transaction abort rates as compared to the other optimistic strategies proposed in literature. In [33], a new **Priority Based Scheme** is proposed, in which the older transactions are getting the first preference whenever conflict arises, thereby decreasing current load of the database server. Assigning priority to transactions for a data item ensures serializability without aborting the transaction.

In [34], a new concurrency control method is proposed **based on AVI value without using locks.** But it has a problem of predicting new AVI value of the data item based on the update history. Ziyad Tariq Abdul-Mehdi et al. develop a mobile transaction model which captures data and movement nature of mobile transaction and propose an innovative concurrency mobile transaction model on **Multi-Check-out Timestamp Order Technique** [35] and present a Serializability theory for this model.

## 4.2. Caching and Query Processing

The main purpose of Caching is to improve data access performance and data availability. Caching data from a central database stored in the Server, performing local queries on the cached data stored in a Mobile device and then finally synchronizing the updated data with the central database has become the common scenario. Caching technique is used to improve the query response time in Client Systems. This technique is very efficient when temporal locality exists in the access patterns between a query and its proceeding queries. Whenever a query is issued, the client cache manager checks its own cache. If it contains a portion of the requested data item then the local query processing starts. To get the remaining portion of the data items, the client sends the request to the server. After the client finishes its local query processing, it stores the result into its cache. Due to the space limitations of the mobile device, it must use a **Cache Replacement Policy** to decide which data item should be replaced when the cache is full. But this technique has a problem of maintaining cache content up-to-date [36].

In [37], Q. Ren and M. H. Dunham use **Quintet structure** by reforming cache item structure, which consists of relations, attributes, predicates, query sender and result receiver. Using this tuple, it is possible to acquire the query constructing this item. It helps to compare new queries with cache items. To improve performance of query processing in mobile databases, a replacement algorithm called FAR is used. This algorithm categorizes the cashed items into two classes such as the items in movement path and the items out of movement path. And then selects the items that are out of movement path to be replaced.

Zhichao Li et al. in [38], define a new structure of semantic segments index. They also analyze the detailed query processing and FAR algorithm. For overcoming the shortage of predicting future location in FAR they propose a new improved algorithm (**RBF-FAR**) as replacement policy. Using this new replacement policy they propose a new model for location-dependent query semantic cache. The experiment results show that new model, on the basis of RBF-FAR, is more flexible and effective at reducing average response time network-load used in LDQ than FAR model.

In [39,40], the semantic cache schema focused on extending domain of query types that can be answered. It consists of cache item structure, cache management algorithm and also cache items replacement algorithm. The proposed schema is based on the concept of **Query Graph Model** (QGM) and all types of queries can be converted to corresponding QGM, hence, the proposed semantic cache schema can answer all types of queries.

Khaleel Mershad et al. propose a system called **S-COACS** which relies on the indexing of cached queries to make the task of locating data more efficient and reliable in MANET [41]. When a new request is issued, nodes cooperate to find its answer and send it to the requesting node. This work semantically compares each request with all cached queries and implements a process that may trim the request into fragments and combine the answers of these fragments to calculate the total answer of the request.

In [42], Bo Yanga and Ali R. Hurson propose a **Semantic-Aware Image Caching** (SAIC) scheme to facilitate content-based image retrieval in wireless **ad-hoc** networks. It can efficiently utilize the cache space and significantly reduce the cost of image retrieval. It is based on several innovative ideas: 1) multilevel representation of the semantic contents, 2) association-based and Bayesian probability-based content prediction, 3) constraint-based representation method showing the semantic similarity between images, 4) nonflooding query processing, and 5) adaptive cache consistency maintenance.

## 4.3. Replication and Synchronization in Mobile databases

Mobile devices do not get continuous connection to the internet and are restricted to limited resources. To improve the availability of data and response time to access the data and to achieve the full offline functionality of data, Replication of the data is the only way to go. Jim Gray et al. in their earlier paper proposed a new Two Tier Replication Algorithm which allows Mobile applications in a disconnected mode to propose tentative update transactions that are later applied to a master copy. The two types of replications used are **Eager and Lazy Replications**. In Eager replication, updates are applied to all replicas as part of one atomic transaction whereas in Lazy replication, only one replica is updated by the original transaction and other replicas need separate transaction for each node. Only Eager replication provides serializable execution without any concurrency anomalies.

Due to the inherent nature of the mobile devices such as intermittent and weak connectivity, restricted bandwidth and inadequate resources an **Optimistic Client Centric Replication Protocol** [43] for relational databases is proposed. This protocol improves the availability by not using the locking mechanisms and reduces the amount of transmitting data between the server and the client by transmitting only the modified data. It also follows an optimistic read-any/write-any approach that provides write access on the client even in the disconnected mode. Only in the client side the primary component of synchronization is placed which uses simple REST interfaces for easy access by devices with limited resources.

To maintain the consistency of the replicated data and to deal with the problems many replication policies are proposed in the literature. These policies are categorized into Pessimistic and Optimistic approaches.

**Pessimistic** strategies restrict updates to a single replica or many replicas by locking access to these replicas during the time of updation. Therefore they offer a strong consistency guarantee that is called one-copy equivalence. They require continuous communication between replicas and blocking of access to these replicas during the operations. But these factors are not feasible in mobile environment due to the intermittent connectivity and less resource availability of the mobile devices. **Optimistic** strategies [44] are under the assumption that inconsistencies rarely occur even if possible. According to this strategy, operations may be executed on any replica. At the time of reconnection, reconciliation will take place during which the replicas exchange all the updates since the last reconciliation. Unlike Pessimistic replication, optimistic replication strategies cannot promise strong consistency but provide weak consistency called eventual consistency.

Ashraf Ahmed et al. in [45] propose a new Optimistic replication strategy for maintaining the eventual consistency in large-scale mobile distributed database systems. It uses the propagation protocol which is used to transfer data updates among the components of replication architecture in such a way that achieves the eventual consistency of data and improves the availability of recent updates to all replicas.

A salient feature of Mobile database systems is to provide disconnected mobile devices the ability to store optimistic replication of data and to perform local updates. But this approach has the reconciliation problem. Two levels of Synchronization techniques found in the literature: **Data Level Synchronization (DLS) and Transaction Level Synchronization (TLS)**. In DLS technique, data is the basic unit of synchronization whereas in TLS transaction is the basic unit. DLS is more simple and highly efficient but it cannot guarantee for Atomicity of transaction and Consistency of the database. TLS guarantees both atomicity and consistency. Therefore TLS is the best suitable technique for synchronization of Mobile database.

In [46], Jose Maria Monteiro et al. propose a new protocol that guarantees one-copy serializability and eventual consistency in mobile computing environments. This protocol uses serializability as correctness criteria, relaxes the isolation property, and adopts the read-any/write-any scheme allowing higher data availability. This approach reduces the amount of data transmission between the servers and the server need not store log.

A new way to implement escrow synchronization without running any type-specific code in the servers, an enhanced method called exo-leasing is presented in [47]. In this method the escrow synchronization code runs in the client and allows a disconnected client to obtain escrow reservation from another disconnected client, reducing the need to coordinate with the servers.
In [48], a reservation mechanism is provided to guarantee that updates can be executed in the server without conflicts. This mechanism provides several types of reservation to guarantee the result of any transaction. Clients obtain leased reservations upon the database state and using this reservation it transparently verifies that a transaction can be executed in the same way both in the mobile client and in the server, thus leading to the same final result. The authors design the **Mobisnap reservation model** to achieve the goal of reservation integrated with mobile transactions in a SQL-based system.

Shirish Hemant Phatak et al., develop a new algorithm [49] that offers **Multiversion reconciliation**. This algorithm integrates both the mechanisms of multiversion concurrency control and reconciliation. The main idea of this algorithm is that the conflict detection is the responsibility of the server while the conflict resolution is the responsibility of the client and they are carried out by global serializability testing. In [50], the authors present Edison, a service that leverages existing off-the-shelf ORDBMS technology to address these problems. It allows large numbers of users to synchronize handheld devices from any point on the internet with subsets of large, shared data sets. It also supports this functionality while transferring the minimal amount of data to and from the device. The implementation of the Edison data server and protocol is described and it is proved that Edison requires minimal overhead in terms of DBMS storage and additional time per synchronization.

In [51], an innovative strategy is proposed that can improve the efficiency of mobile database and the success ratio of committing transient transaction. According to this strategy, the data in the Mobile Client (MC) are categorized into 3 groups: private data, public data and shared data depends on the properties of the data. It suggests different data synchronizing strategy for different data and analyses the data synchronizing mechanism under three state of MC cache: consistency, disconnection and integration.

A new model of transaction synchronization and replication is proposed in [52] to overcome the defects existing in current technology of synchronizing replication for mobile databases. In this model, the conflict preprocessing mechanism based on user-concerned data and transaction-related-set is adopted and executed before conflict detection and resolution. The conflict reconciliation strategy based on rule simplifies conflict processing and enhances the efficiency of synchronization and replication.

The synchronization and conflict resolution for Transport Canada's RSIG (Rail Safety Integrated Gateway) system is presented in [53]. The authors provide a complete solution to address the challenges of a Disconnected Mode problem in the RSIG system based on Microsoft SQL Server database technology using merge replication. Merge replication is used for mobile or distributed server applications with possible data conflicts. The advantage of this solution is that it can be used by retail wholesalers.

In [54], the authors propose Split Synchronization Mobile Transaction (SSMT) Model that assures atomicity in synchronization, reduces synchronization cost and provides context-awareness. This model divides the synchronization process of a mobile transaction into 2 parts namely preferred and deferred parts by exploiting the hoardset fragmentability of data objects which the mobile transaction access. The preferred synchronization part with possibility of conflicts is synchronized early using the transaction-centric synchronization strategy whereas the deferred synchronization part which has no possibility of conflict uses the **data-centric** synchronization strategy and its synchronization is delayed.

 **Data Centric synchronization** approach is used based on the Publish/Subscribe Model [55, 56, 57]. In this model, the server should know the caching requirements of all mobile clients in advance and predefine publications satisfying them and publish the publication. The Client (Mobile Client) should subscribe to publications and the snapshots of data items in the subscribed publications are created as its mobile database. Then the client processes transaction with the snapshots in a disconnected mode and finally connects to the server and synchronizes the snapshots with the server database to maintain the cache coherency.

A tiered replication model [11] is used to solve the **Transactional** based synchronization problem under 3-tier replication architecture. To promote the efficiency of synchronization, it employs a new Double Timestamp strategy. In this model each tuple in the mobile database has 2 Timestamps called Primary and Secondary Timestamps. While modifying the tuple, the time mark added by the server is the primary while the transaction mark added by the mobile database is the secondary timestamp. It also ensures serializability and consistency of the mobile database system through **transaction-level** conflict reconciliation.

## 4.4. Recovery in Mobile Databases

In Mobile Database System Recovery is highly challenging since some of the processing nodes are mobile. Other challenges in the recovery process include, failures at the fixed network DBMS, limited wireless channels, network failure, site failure and handoff problems. Solutions to these challenges weaken atomicity or isolation levels of transactions. Two types of protocols namely **Checkpoint and Message Logging** are used for saving the execution state of a mobile application, so that when a MH recovers from a failure, the mobile application can roll back to the last saved consistent state, and restart execution with recovery guarantees. Checkpoint and log information are stored at the base stations [57] since Mobile Host (MH) disk storage is not stable.

### 4.4.1. Checkpoint Protocols

Efforts are taken to find an apt protocol that could perform well in the mobile environment despite the limitations of it.  In the literature three kinds of approaches are found: Uncoordinated Checkpoint, Coordinated Checkpoint and Communication-Induced Checkpoint.

In a distributed environment, if each participating process takes its checkpoints independently, then the system is prone to the domino effect (cascaded rollback). This approach is called **Uncoordinated Checkpointing** [58]. If the processes coordinate their checkpoints to form a system-wide consistent state, then the approach is called **Coordinated Checkpointing** [59]. In this approach, if there is any failure, the system state can be restored to a consistent set of checkpoints and preventing the rollback propagation. If every process is forced to take checkpoints based on information piggybacked on the application messages it receives from other processes then this approach is called **Communication-Induced Checkpointing.** Checkpoints are taken such that a system-wide consistent state always exists on stable storage, thereby avoiding the domino effect.

One popular recovery protocol is **Timeout protocol**, executed by MSS which maintains a timer to measure the inactivity period of mobile host and initiates rollback for the transaction on timeout [60]. In [61], **Mutable Checkpoint** to design efficient checkpointing algorithm is introduced. It can be saved anywhere (the memory or local disk of MHSs). Taking a mutable checkpoint avoids the overhead of transferring large amount of data to stable storage at MSSs over the wireless network. The proposed algorithm based on the Mutable Checkpoint is a **non-blocking algorithm** that avoids the avalanche effect and forces only a minimum number of processes to take their checkpoints on the stable storage.

Bidyut Gupta et al. present a **non-blocking coordinated checkpointing** algorithm [62] suitable for mobile environments. In this proposed algorithm the overhead of converting temporary checkpoint to permanent checkpoint and mutable checkpoint to permanent ones are eliminated. Also it uses very few control messages and participating processes are interrupted less number of times. The algorithm produces a consistent global state of the system.

In [63], Awadhesh Kumar Singh et al. present **index-based checkpointing** and recovery protocol. As this protocol takes only fewer checkpoints and does not require computing dependency relationships, it is more efficient in computation. It is nonblocking, adaptive and uses few control messages. Irrespective of the message sending rate, it works well in all situations. In [64], Rachit Garg et al. present a survey of some checkpointing algorithms for distributed systems.

In [65], Sungchae Lim proposes a **checkpointing algorithm** that allows the distributed mobile application to tune the level of its checkpointing strictness (the maximum rollback distance) so that the usage of wireless communications can be reduced.    Using this algorithm there is a possibility of reducing the number of enforced local checkpointing. The advantage of this algorithm is less communication cost and shortened blocking time.

In [66], Cheng-min lin et al. provide a **domino effect-free failure recovery algorithm** which contains three-phases that ensures a consistent checkpoint. In the first phase, it uses a coordinated checkpointing protocol among mobile support stations; in the second phase, a communication-induced checkpointing protocol is used between each MSS and its MHs and in the last phase, each MSS sends a checkpoint request to its mobile host which hasn't received any message from the mobile support station during the second phase. Domino effect-free, nonblocking, twice the checkpoint size, and scalability are the advantages of this algorithm.

In [67], Sapna E. George et al. present an efficient failure recovery scheme for mobile database applications based on **movement-based checkpointing** and logging. It combines independent checkpointing and pessimistic message logging enabling asynchronous recovery of a MH upon failure. In this scheme, checkpoint is taken only after a threshold of mobility handoffs has been exceeded. When the MH crosses a distance threshold from the location of the latest checkpoint, the recovery information is collected and transferred to the MH's local MSS.   The authors identify the optimal movement threshold which will minimize the recovery cost per failure as a function of the mobile node's mobility rate, failure rate and log arrival rate.

In [68], Chandreyee Chowdhury et al. present a **coordinated non blocking checkpointing** and recovery technique for such systems that handles well the constraints posed by the underlying wireless network. In this technique, an initiator (an MSS) sends checkpoint requests to all other MSSs and the MSSs send this request only to those MHs, which have communicated in the last checkpointing interval and thereby relieving the wireless network from synchronization overhead. Also all acknowledged messages are logged at the home station of the receiver MH so that only the faulty MHs need to recover in case of failure and no other process is affected by this fault and subsequent recovery.

In [69], S. Kumar et al. provide the concept of **hybrid checkpointing** algorithm, where in an all-process coordinated checkpoint is taken after the execution of minimum process coordinated checkpointing algorithm for a fixed number of times.

In [70], Biswas et al. propose a checkpointing and failure recovery algorithm where MHs save checkpoints based on mobility and movement patterns. Mobile hosts save checkpoints when number of hand-offs exceed a predefined handoff threshold value. They introduced the concept of migration checkpoint. An MH upon saving migration checkpoint sends it attached with migration message to its current MSS before disconnection.

Ruchi Tuli et al. review the modern checkpointing schemes in distributed and mobile environments and present a comparative study of the protocols or schemes based on their relative advantages and disadvantages in [71].

### 4.4.2. Message Logging Protocols

Ing-Ray Chen et al. report **Lazy and Pessimistic** schemes in [72]. In a lazy scheme, logs are stored in the Base station (BS) and if mobile host moves to a new BS, a pointer to the old BS gets stored in the new BS. Pointers can be used to recover the log distributed over several BS's. This scheme has the advantage that it incurs relatively less network overhead during handoff as no log information needs to be transferred. But this scheme has a large recovery time. In the pessimistic scheme, the entire log and checkpoint record, if any, are transferred at each handoff. Hence, the recovery is fast but each handoff requires large volume of data transfer.

In [73], schemes based on the mobile host's movement using independent check pointing and **pessimistic logging** are given. In the distance-based scheme, the contents that are distributed are unified when the distance covered by mobile host increases above a predefined value. After unifying the log, the distance or handoff counter is reset. These schemes are a tradeoff between lazy and the pessimistic strategies. The schemes discussed so far do not consider the case where a mobile host recovers in a base station different from the one in which it has crashed. Jiang et al. [74] present an **optimistic checkpointing and selective message logging approach** for consistent global checkpoint collection in distributed systems.

Sashidhar Gadiraju et al. in [75] present an **application log management scheme**, which uses a mobile-agent-based framework to facilitate seamless logging of application activities for recovery from transaction or system failure. They present the forward strategy for improving the recovery time in situations where the failure time is nontrivial. The framework is not restricted to the forward scheme, but it can support previous independent logging schemes. The simulation results show that the forward scheme improves the recovery time with a fairly consistent behavior in all the parameters simulated.

J.C. Miraclin Joyce Pamila et al. in [76] propose a **new log management scheme** for recovery of mobile transactions. In this approach, the model parameters that affect application state recovery are analyzed. It reduces total cost for recovery from the failure when compared with the existing Lazy and pessimistic schemes. It also ensures recovery from different BS other than in which it has failed. The proposed scheme controls the handoff cost, log retrieval cost and failure recovery time

## 4.5. Security in Mobile databases

In his previous paper, Astrid Lubinski presented mobility related security issues and categorizes the problems by risk for data and metadata in their access, management and transfer. He identifies the protection of user whereabouts and movements as a main risk in mobile environment and suggests 3 separations namely Horizontal, Vertical and Aggregate separation in Access and management to guard this additional information. He addresses an Adaptation component in [77] that is used to mediate all accesses and to manage security relevant context modification. M.A. Badamas provides minimal security measures and discusses the implementation of access control method based on application at the beginning of the hard disk and the role of encryption in securing the data [78].

In [79], the authors propose a new Client-based security component called Chip-Secured Data Access (C-SDA) that enforces security requirements such as Confidentiality and Authentication in the client. It is fixed into the smart card to prevent any kind of tampering to occur on the client side. This combination of hardware and software security component provides guarantee against attacks and threatening personal as well as business data. However, database attacks are more and more frequent (their cost is estimated to be more than $100 billion per year) and 45% of the attacks are conducted by insiders [80]. Several attempts have been made to strengthen server-based security approaches with database encryption [81]. However, as Oracle confesses, server encryption is not the expected "armor plating" because the Database Administrator (or an intruder usurping her identity) has enough privilege to tamper the encryption mechanism and get the clear-text data.

In [82], the authors discuss the security policy of mobile databases and present the various safety measurements to solve the problems. To ensure the safety of mobile database they suggest few mechanisms such as authentication, encryption, auditing, backup and recovery. Parviz et al., present threats and solutions for Mobile databases in [83]. They discuss that in order to secure Mobile Databases, Mobile device, Operating system on mobile device, and Mobile network are also secured. In [84], the author proposes a Paradigm for securing Mobile Databases and identifies three areas to be secured such as Client side security, Server side security and Security during data transmission. She also provides the advantages of the proposed system.

Even though many papers have been found in the literature on security policies, security mechanisms, problems and solutions for mobile databases, none of them provide a comprehensive solution and practical implementation for securing the mobile databases to its entirety. The recent papers [82, 83] on security recommend that there is a need to continue to explore the security issues of database system in the mobile environment.

Research works related to individual entity such as Mobile Device security, Client side security, Server side security and Security during data transmission are going on. But these topics are not focused in this paper due to space limitations.

## 4.6. Issues and Research Directions

In [85], Grenoble et al. address the various issues in mobile database systems as Copy Synchronization in disconnected computing, Mobile Transactions, Database embedded in Ultra-light Devices, Data Confidentiality, P2P dissemination models and Middleware Adaptability. In [2], Weider Yu et al. categorize the issues and solutions for designing the mobile database as Responsiveness, Data consistency and Concurrency, Synchronization and Conflict resolution, Security, High Availability, Data size, Screen Size, Slow Transmission Speed, Slow Processing Speed and Cost. In [9, 86] the authors discuss the important issues of Mobile databases as the relative unreliability of connections (and the variability of bandwidth when connected), the limitations on storage capacity and the security and privacy issues created when a computer is in a mobile environment. After reviewing the issues of the mobile databases from many literatures, it is found that all the authors have invariably addressed the Security and Privacy as one of the important critical issues and then the limitations of storage capacity and unreliable connectivity. Table 1 shows the various issues given in the corresponding papers.

| Papers \ Issues | [GRE04] | [WEI08] | [DAR05] | [DAR03] | [SRI09] |
|---|---|---|---|---|---|
| Mobile Transaction | ✓ | ✓ | | | |
| Synchronization | ✓ | ✓ | | | |
| Data Consistency | ✓ | ✓ | | | |
| Concurrency | ✓ | ✓ | | | |
| Security and Privacy | ✓ | ✓ | ✓ | ✓ | ✓ |
| Database embedded in Ultra Light device | ✓ | | | | |
| P2P Dissemination Model | ✓ | | | | |
| Limitations of Storage Capacity | | ✓ | ✓ | ✓ | ✓ |
| Unreliable Connectivity | | | ✓ | ✓ | ✓ |
| Performance | | ✓ | | | |

Table 1 Papers Vs Issues

## 5. CONCLUSION

In recent past, Industry and Academia has shown a lot of interest in improving data and transaction management in Mobile environment. Data and Transaction management in mobile database is very difficult compare to traditional databases as the mobile environment needs to be highly versatile and have to satisfy several resource constraints. Hence this field attracts a lot of research in terms of designing new algorithms, techniques or strategies for the various resource constraints on the functionalities of Mobile Database including Security.

# REFERENCES

[1] Ouri Wolfson, "Mobile Database", Encyclopedia of Database Systems, Part 13, Page 1751, 2009. http://www.springerlink.com/content/n72wu51n4056524g/fulltext.html, Springer Science+Business Media, LLC 2009, 10.1007/978-0-387-39940-9_1362.

[2] Weider Yu, D., Tamseela Amjad, Himani Goel and Tanakom Talawat, "An Approach of Mobile Database Design methodology for Mobile Software Solutions", The 3rd International Conference on Grid and Pervasive Computing - Workshops, IEEE 2008.

[3] S.K. Singh, "Database Systems Concepts, Designs and Applications", Dorling Kindersley (India) Pvt. Ltd., New Delhi, 2006.

[4] Thomas Connolly, Carolyn Begg, "Database Systems: A Practical Approach to Design, Implementation and Management", 3rd ed., Dorling Kindersley (India) Pvt. Ltd., New Delhi, 2002.

[5] Fuentes, E., Knapp, L., Burke, D., Introduction to Mobile Databases. www.cin.ufpe.br/~fcdf/Middleware/projeto/Referencias/Mobile_Databases.ppt

[6] Weibo Li, Hong Yang and Ping He, "The Research and Application of Embedded Mobile Database", IEEE, Computer Society, 2009.

[7] Vijay Kumar, "Mobile Database Systems", Wiley-Interscience, John Wiley & Sons, Inc. Publication, New Jersey, 2006.

[8] Yanli Xia and Abdelsalam Helal, "A Dynamic Data/Currency Protocol for Mobile Database Design and Reconfiguration", SAC 2003, ACM.

[9] Darin Chan and John Roddick, F., "Summarisation for Mobile databases", Journal of Research and Practice in Information Technology", Vol. 37, No.3, 267—284, August 2005.

[10] Yang Li, Xuejie Zhang, Yun Gao, "Object-Oriented Data Synchronization for Mobile Database over Mobile Ad-hoc Networks, International Symposium on Information Science and Engieering", IEEE, 2008.

[11] Zhnag Yun , Zhang Mu, Bian Fuling, "A Tiered Replication Model in Embedded Database Based Mobile Geospatial Information Service", IEEE, 2008.

[12] Jing Li, Jaianhua Wang: A New Architecture Model of Mobile Database Based on Agent, First International workshop on Database Technology and Applications, IEEE, 2009.

[13] S.K. Madria and B. Bhargava, "A transaction model for improving data availability in mobile computing", Kluwer Academic Publishers Distributed and Parallel Databases, Vol. 10, No. 2, 2001.

[14] M. Lee and S. Helal, "HiCoMo: High Commit Mobile Transactions," Kluwer Academic Publishers Distributed and Parallel Databases (DAPD), Vol. 11, No. 1, 2002.

[15] K. Ku and Y. Kim, "Moflex transaction model for mobile heterogeneous multidatabase systems," IEEE Workshop on Research Issues in Data Engineering, San Diego, USA, Feb. 2000.

[16] Ravi A. Dirckze and Le Gruenwald, "A Pre-serialization transaction management technique for mobile multidatabases," Mobile Networks and Applications (MONET), Vol. 5, No. 4, 311–321, 2000.

[17] Kam-Yiu Lam, GuoHui Li and Tei-Wei Kuo, "A Multi-Version Data Model for Executing Real-time Transactions in a Mobile Environment", MobiDE 2001,ACM, 2001.

[18] P. Serrano-Alvarado, C. Roncancio, and M. Adiba, "A Survey of Mobile Transactions", Kluwer Academic Publishers Distributed and Parallel Databases(DAPD), 16(2), 2004.

[19] Joos-Hendrik Bose et al., "Research Issues in Mobile Transactions", Proceedings of Dagstuhl Seminar on Mobile Information Management, 2004.

[20] Prasanta Kumar Panda Sujata Swain P. K. Pattnaik, "Review of Some Transaction Models used in Mobile Databases", International Journal of Instrumentation, Control & Automation (IJICA), Volume 1, Issue 1, 2011. Computing (SCC'04), IEEE Computer Society, 2004.

[21] Simon Cuce, Arkady Zaslavsky, Bing Hu and Jignesh Rambhia, "Maintaining Consistency of Twin Transaction Model Using Mobility-Enabled Distributed File System Environment", Proceedings of the 13th International Workshop on Database and Expert Systems Applications (DEXA'02), IEEE, 2004.

[22] B. Zheng, "On Semantic Caching and Query Scheduling for Mobile Nearest-Neighbor Search", ACM Wireless Network Conference, Hingham, MA, USA, November 2004.

[23] Patricia Serrano-Alvarado, Claudia Roncancio, Michel Adiba, Cyril Labbe, "An Adaptable mobile transaction model", International conference on Computer Systems Science and Engineering, 2005.

[24] Osman Mohammed Hegazy, Ali Hamed El Bastawissy, Romani Farid Ibrahim, "Handling Mobile Transactions with Disconnections Using a Mobile-Shadow Technique", INFOS2008, Cairo-Egypt, March 27-29, 2008.

[25] Ravimaran. S, Maluk Mohamed. M.A., "An Improved Kangaroo Transaction Model Using Surrogate Objects for Distributed Mobile System", MobiDE'11, ACM, 2011.

[26] Tome Dimovski, Pece Mitrevski, "Connection Fault-Tolerant Model for Distributed Transaction Processing in Mobile Computing Environment", Proceedings of the ITI 33rd   Int. Conf. on Information Technology Interfaces, June 27-30, 2011, Cavtat, Croatia.

[27] Turker, C., Zini, G., "A Survey of Academic & Commercial Approaches to Transaction Support in Mobile Computing Environments", 2003.

[28] Vijay Kumar, Nitin Prabhu, Maggie Dunham, Ayse Yasemin Seydim, "TCOT: A Timeout based Mobile Transaction Commitment Protocol", IIS 9979453, 2004.

[29] Salman Abdul Moiz, Dr. Lakshmi Rajamani, "Single Lock Manager Approach for achieving Concurrency in Mobile Environments", 14th   IEEE International Conference on High Performance Computing (HiPC), 2007. Springer LNCS 4873, ISBN 978-3-540-77219-4, 650-660, 2007.

[30] Salman Abdul Moiz, Dr. Lakshmi Rajamani, "An Efficient Strategy for achieving Concurrency Control in Mobile Environments", 12th IEEE Asia Pacific Network Operations & Management (APNOMS) symposium, Springer LNCS 5787, ISBN# 978-3-642-04491-5, 519-522, 2009.

[31] Victor C.S.Lee, Kwok Wa Lam, Tei-wei Kuo,"Efficient Validation of Mobile Transactions in Wireless Environments", The Journal of Systems and Software 69, 183-193, 2004.

[32] Salman Abdul Moiz and Dr. Lakshmi Rajamani,   "A Real Time Optimistic strategy to achieve Concurrency Control in Mobile Environments using  On-demand Multicasting", International Journal of Wireless and Mobile Networks, Vol, 2, No. 2, 2010.

[33] Mohammed Khaja Nizamuddin and Syed Abdul Sattar, "Algorithm for Priority Based Concurrency Control without Locking in Mobile Environments", IEEE, 2011.

[34] Joe, C.H.Y., Chan, E., Lam, K.Y., Leung, H.W, "An Adaptive AVI-based Cache Invalidation Scheme for Mobile Computing Systems", DEXA 2000, pp. 155-178, IEEE Computer Society, USA.

[35] Ziyad Tariq Abdul-Mehdi, Ali Bin Mamat & Hamidah Ibrahim, Mustafa M. Dirs , "New Concurrency Model for Mobile Database Environment", International Journal of Theoretical and Applied Computer Sciences, GBS Publishers and Distributors, Vol. 1, No. 1, 101–108, 2006.

[36] Yoshiharu Ishikawa, Hiroyuki Kitagawa, "A Semantic Caching Method Based on Linear Constraints", IEEE, 2000.

[37] Q.Ren, M.H.Dunham," Semantic Caching in Mobile Computing", PhD thesis, Southern Methodist University, Computer Science and Engineering, Dallas, TX 75205, February 2000.

[38] Zhichao Li, Pilian He, Ming Lei, "Research of Semantic Caching for LDQ in Mobile Network", IEEE, 2005.

[39]  A. Guy M. Lohman, "the DB2 Universal Database Optimizer", IBM Research Division , IBM Almaden Research Center, 2007.

[40]  Ali Safaeei, Mostafa Haghjoo, Sulmaz Abdi, "Semantic Cache Schema for Query Processing in Mobile Databases", IEEE, 2008.

[41] Khaleel Mershad, Hassan Artail, "Semantic Caching for Mobile Ad Hoc Networks", Fifth International Conference on Mobile Ad-hoc and Sensor Networks, IEEE, 2009.

[42] Bo Yanga and Ali R. Hurson, "Semantic-Aware and QoS-Aware Image Caching in Ad Hoc Networks", IEEE Transactions On Knowledge And Data Engineering, Vol. 19, No. 12, December 2007.

[43] Martin Peters, Christopher Brink, Martin Hirsch, Sabine Sachweh, "A Client Centric Replication Model for Mobile Environments based on RESTful Resources", ACM, 2011.

[44] Y. Saito and M. Shapiro, "Optimistic replication", ACM Computing Surveys, Vol. 73, No. 1, 42 – 81, 2005.

[45] Ashraf Ahmed, P.D.D.Dominic, Azween Abdullah and Hamidah Ibrahim, "A New Optimistic Replication Strategy for Large-scale Mobile Distributed Database Systems", International Journal of Database Management Systems, Vol.2, No.4, 86-105, Nov. 2010.

[46] Jose Maria Monteiro, Angelo Brayner, Sergio Lifschitz, "A Mechanism for replicated data consistency in mobile computing environments", Proceedings of the ACM symposium on Applied computing, Seoul, Korea, 914 – 919, 2007.

[47] Liuba Shrira, Hong Tian, and Doug Terry, "Exo-Leasing: Escrow Synchronization for Mobile Clients of Commodity Storage Servers", IFIP International Federation for Information Processing, V. Issarny and R. Schantz (Eds.): Middleware 2008, LNCS 5346, 42–61, 2008.

[48] Nuno Preguica et al., "Reservations for Conflict Avoidance in a Mobile Database System", First International Conference on Mobile Systems, Applications, and Services, 2003.

[49] Shirish Hemant Phatak and Badri Nath, "Transaction-Centric Reconciliation in Disconnected Client–Server Databases", Mobile Networks and Applications 9, Kluwer Academic Publishers, Netherlands, 459–471, 2004.

[50] Matthew Denny and Michael J. Franklin, "Edison: Database-Supported Synchronization for PDAs", Kluwer Academic Publishers, Netherlands, Distributed and Parallel Databases, 15, 95–116, 2004.

[51] Wang Shun Yan, Zhong Luo, Cao Yong Liang, "A Data Synchronization Mechanism for Cache on Mobile Client", IEEE, 2006.

[52] Min Jin, Xiang Zhou, Jihui Zhou, Xianming Gao and Chunhong Gong, "Strategy of Conflict Preprocessing and Reconciliation for Mobile Databases", IEEE , 2008.

[53] Samuel A. Ajila, Ahmed Al-Asaad, "Mobile Databases - Synchronization & Conflict Resolution Strategies using SQL Server", IEEE, 2011.

[54] Miseon Choi , Wonik Park and Young-Kuk Kim, "A Split Synchronizing Mobile Transaction Model", Proc. of the 2nd Int. National Conference on Ubiquitous Information Management and Communication, New York, USA, 196 – 201, 2008.

[55] Oracle 10g Documentation, http:// www.oracle.com

[56] Sybase UltraLite Online Documentation, http:// www.sybase.com

[57] S. Gadiraju, Vijay Kumar, "Recovery in the mobile wireless environments using mobile agents", IEEE Transactions on Mobile Computing, Vol. 3, June 2004.

[58] Ni, W., Vrbsky, S.V., Ray, S., "Low-cost Coordinated Checkpointing in Mobile Computing Systems", Proceeding of the Eighth IEEE International Symposium on Computers and Communications, 2003.

[59] Elonzahy, E.N., Alvisi, L., Wang, Y.M., Johnson, D.B., "A Survey of Rollback-Recovery protocols in Message-Passing Systems", ACM Computing Surveys, Vol.34, No. 3, 375–408, 2002.

[60] M. M. Gore , R. K. Ghosh, "Recovery of Mobile Transactions", Proc. of the 11th International Workshop on Database and Expert Systems Applications, 2000.

[61] G. Cao and M. Singhal, "Mutable Checkpoints : A New Checkpointing Approach for Mobile Computing Systems", IEEE Transactions On Parallel And Distributed Systems,Vol.12, No.2, 157-172., February 2001.

[62] Bidyut Gupta, S. Rahimi and Z. Lui, "A New High Performance Checkpointing Approach for Mobile Computing Systems". IJCSNS International Journal of Computer Science and Network Security, Vol.6, No.5B, May 2006.

[63] Awadhesh Kumar Singh, Rohit Bhat, and Anshul Kumar, "An Index-Based Mobile  checkpointing and Recovery Algorithm", ICDCN 2009, LNCS 5408, 200–205, Springer-Verlag Berlin Heidelberg, 2009.

[64] Rachit Garg, Praveen Kumar, "Low Overhead Checkpointing Protocols for Mobile Distributed Systems: A Comparative Study", International Journal of Engineering Science and Technology Vol. 2, No. 7, 3267-3276, 2010.

[65] Sungchae Lim, "A Tunable Checkpointing Algorithm for the Distributed Mobile Environment", IJCSI International Journal of Computer Science Issues, ISSN (Online): 1694-0814, Vol. 8, Issue 6, No 3, November 2011.

[66] Cheng-min lin and chyi-ren dow, "Efficient checkpoint-based failure recovery techniques in Mobile computing systems", Journal of information science and engineering 17, 549-573, 2001.

[67] Sapna E. George and Ing-Ray Chen , "Movement-based checkpointing and logging for failure recovery of database applications in mobile environments", Distrib Parallel Databases 23, Springer Science+Business Media, 189–205, 2008.

[68] Chandreyee Chowdhury and Sarmistha Neogy, "A Consistent Checkpointing-Recovery Protocol for Minimal Number of Nodes in Mobile Computing System", S. Aluru et al. (Eds.), Springer-Verlag Berlin Heidelberg, LNCS 4873, 599–611, 2007.

[69] Kumar, P., Garg, R., "Soft Checkpointing Based Hybrid Synchronous Checkpointing Protocol for Mobile Distributed Systems", International Journal of Distributed Systems and Technologies, Vol. 2, No.1, 1–13, 2011.

[70] Biswas, S., Neogy, S., "A Mobility-Based Checkpointing Protocol for Mobile Computing System". International Journal of Computer Science & Information Technology, Vol. 2, No.1, 135–151, 2010.

[71] Ruchi Tuli and Parveen Kumar, "New Paradigms in Checkpoint Processing and Recovery Techniques for Distributed Mobile Systems", D.C. Wyld et al. (Eds.), NeCoM/WeST/WiMoN 2011, CCIS 197, Springer-Verlag Berlin Heidelberg, 221–231, 2011.

[72] Ing-Ray Chen, Baoshan Gu, Sapna E. George, and Sheng-Tzong Cheng, "On Failure Recoverability of Client Server Application in Mobile Wireless Environments", IEEE Transactions on Reliability, Vol. 54, March 2005.

[73] George, S.E., Chen, I.-R., Jin, Y, "Movement Based Checkpointing and Logging for Recovery in Mobile Computing Systems", ACM, New York, 2006.

[74] Jiang, Q., Manivannan, D., "An Optimistic Checkpointing and selective message logging approach for consistent global checkpoint collection in distributed systems", IEEE, Los Alamitos, 2007.

[75] Sashidhar Gadiraju and Vijay Kumar, "Recovery in the Mobile Wireless Environment Using Mobile Agents", IEEE Transactions on Mobile Computing, Vol. 3, No. 2, June 2004.

[76] J.C. Miraclin Joyce Pamila, K. Thanushkodi, "Log Management Support for Recovery in Mobile Computing Environment", International Journal of Computer Science and Information Security, Vol. 3, No. 1, 2009.

[77] Astrid Lubinski, "Adaptation Concepts for Mobile Database Security", Proc. of the IEEE International Symposium on Database Technology and Software Engineering, 445-450, 2000.

[78] M.A. Badamas, "Mobile Computer Systems - Security Considerations", Information Management & Computer Security, MCB University Press, 2001, 134 – 136.

[79] Luc Bouganim, Philippe Pucheral, "Chip-Secured Data Access: Confidential Data on Untrusted Servers", Proceedings of the 28th VLDB Conference, Hong Kong, China, 2002.

[80] Computer Security Institute, "CSI/FBI Computer Crime and Security Survey", http://www.gocsi.com/forms/fbi/pdf.html

[81] Oracle Corp., "Advanced Security Administrator's Guide", Release2, (9.2), 2002.

[82] Huaixiang Wang DeyuDang Shi Min, "The Analysis of the Security Strategy of Embedded Mobile Database", IEEE, 2010.

[83] Parviz Ghorbanzadeh, Aytak shaddeli, Roghieh Malekzadeh, Zoleikha Jahanbakhsh, "A Survey of Mobile Database Security Threats and Solutions for It", 3rd International Conference on Information Sciences and Interaction Sciences (ICIS), 676-682, China, 2010.

[84] Roselin Selvarani .D, "A Secured Paradigm for Mobile Databases", N. Meghanathan et al. (Eds.), Proceedings of 3rd International Conference on Recent Trends in Network Security and Applications (CNSA 2010), Springer Communications in Computer and Information Science 89, ISBN 978-3-642-14477-6, ISSN 1865-0929, Springer-Verlag Berlin Heidelberg, 164–171, 2010.

[85] EPFL, Grenoble, U., INRIA-Nancy, INT-Evry, Montpellier, U., Paris, U. Versailles, U., "Mobile Databases: a Report on Open Issues and Research Directions", SIGMOD Record, Vol. 33, No. 2, June 2004.

[86] A.Sripriya, Dr. R.Dhanapal, "A Cache Operation in Mobile Database Using Genetic Algorithm", International Journal of Recent Trends in Engineering, Vol 2, No. 4, November 2009.